\documentclass[prc,nofootinbib]{revtex4-1}
\usepackage{amsmath,amssymb,bm}
\newcommand{\braket}[1]{\langle {#1} \rangle }
\newcommand{\ket}[1]{|{#1} \rangle }
\newcommand{\bra}[1]{\langle {#1}|}

\usepackage{amsmath}	
\usepackage{color}
\usepackage[pdftex]{graphicx}
\usepackage{hyperref}

\begin{document}
\title{Merging ab initio theory and few-body approach for  $(d,p)$ reactions}

	\author{J. Rotureau}
        \affiliation{NSCL/FRIB Laboratory, Michigan State University, East
          Lansing, Michigan 48824, USA}
        
	\author{G. Potel}
        \affiliation{NSCL/FRIB Laboratory, Michigan State University, East
          Lansing, Michigan 48824, USA}
     
	\author{W. Li}
        \affiliation{NSCL/FRIB Laboratory, Michigan State University, East
          Lansing, Michigan 48824, USA}

	\author{F.M. Nunes}
        \affiliation{NSCL/FRIB Laboratory, Michigan State University, East
  Lansing, Michigan 48824, USA}

	\date{\today}
	\begin{abstract}
          A framework for $A(d,p)B$ reactions is introduced by merging the microscopic approach
          to computing the properties of the nucleon-target systems  and  the three-body $n+p+A$ reaction formalism, thus providing a consistent link between the reaction cross sections  and the underlying microscopic structure. In this first step toward a full microscopic
          description, we focus on the inclusion of the neutron-target microscopic properties.
          The properties of the neutron-target subsystem are encapsulated in the Green's function which is 
          computed with  the Coupled Cluster theory using a chiral nucleon-nucleon and three-nucleon interactions. Subsequently, this many-body information is introduced in the few-body Green's Function Transfer approach to $(d,p)$ reactions.
Our benchmarks  on stable targets $^{40,48}$Ca show an excellent agreement with the data.
We then proceed to make specific predictions for $(d,p)$ on  neutron rich $^{52,54}$Ca isotopes. These predictions are directly relevant to testing the new magic numbers $N=32,34$ and are expected to be feasible in the first campaign of the projected FRIB facility. 
\end{abstract}	

	\maketitle

        \section{Introduction}
 Progress on the capability to produce rare--isotopes beams (RIBs \cite{fac,frib_lab,fair}) has  pushed the exploration frontier into remote parts of the nuclear chart far from the valley of stability. The expectation that our traditional knowledge would be challenged as one treads through these exotic nuclear regions has been experimentally confirmed. A striking example is provided by the emergence of new magic numbers, {\it i.e} the number of nucleons  that fill major shells. Magic numbers are one of the cornerstones of nuclear structure, and nuclei with magic numbers of protons and/or neutrons display a larger stability compared to their close neighbors. A recent example is the experimental evidence of new doubly--magic features
in the short--lived $^{52,54}$Ca  \cite{wienholtz2013,steppenbeck2013b,michimasa}.

Nuclear reactions play a key role in the experimental study of nuclei, offering a variety of probes allowing to extract
complementary information about the structure of the systems under study. Within this context, one-nucleon transfer reactions such as $(d,p)$ are the probe of choice to obtain information about the nuclear response to nucleon addition (single-particle strength)   as a function of energy, angular momentum and parity.
By comparing experimental data to theoretical predictions, reaction cross sections  can also be used as a tool 
to inform, validate and refine theoretical structure models. But, in order to extract unambiguous  information from reaction observables, it is essential to integrate   consistently the structure theory in the reaction formalism. This is the main objective of this paper.

Although some recent works  describe $(d,p)$ reactions {\it ab-initio} \cite{Raimondi:16,hupin}, for most cases of interest one usually relies on the reduction of the many--body problem to a few--body one where only the most relevant
degrees of freedom are retained \cite{Satchler:83,Thompson:09}.
In this picture, the Hamiltonian is given
as a sum of two-body effective interactions 
between the clusters considered.  A standard approach  to obtain the two-body interactions is to fit a simple local function (e.g., a Woods-Saxon) from experimental elastic scattering data  on $\beta$--stable isotopes \cite{KD,pheno}. As no explicit connection to an
underlying microscopic theory is made,
these potentials become less reliable and bring uncontrolled uncertainties as one considers systems further from stability.
In the most common approaches in the field   \cite{Satchler:83,Thompson:09}   the $(d,p)$ cross section is factorized into a single-particle  reaction term (which accounts for the dynamics of the process) and a spectroscopic factor (which relates to the probability of a certain orbital configuration in the final state). Unlike cross sections, spectroscopic factors and potentials
are non-observable quantities  \cite{Mukha,Furnstahl_2010,Jennings:2011jm,Duguet:15}. They depend on the  model and the representation used to
compute them
\footnote{Non-observable quantities like potentials and spectroscopic factors are not uniquely defined. Spectroscopic factors
 are not invariant under finite-range unitary transformations which translates into a dependence on the model space
 and the interactions used to compute them \cite{Mukha,Furnstahl_2010,Jennings:2011jm,Duguet:15}.   
 For a given  potential, it is possible to modify its high-energy component with a unitary transformation
 without affecting experimental predictions \cite{bogner09}.}.
Unless  they are  calculated consistently  within a  same framework,
serious calibration issues in the theory could appear.
This is likely to become even more problematic when moving toward uncharted territories of the nuclear landscape.
One must then strive to compute all inputs of the few-body problem consistently.

In this paper, we introduce a framework that combines the  development in obtaining the microscopic effective interactions from coupled cluster (CC) theory \cite{gfccpap,gfccpap2}  and the Green's Function Transfer (GFT) reaction theory for $(d,p)$ on medium-mass nuclei \cite{Potel:15b,Potel:17b}. In this merged framework, the structure content of the target and the neutron-target effective interactions
are consistently computed with the CC approach, resulting from the same underlying many-body Hamiltonian.
The CC Green's Function and neutron-target potential are then susbsequently integrated in the GFT few-body formalism.
Other inputs entering the  GFT equations  are a $p-(A+1)$ and  a $d-A$ optical potentials (see Sec.~\ref{formalism}).
In this first application, these potentials will be taken from phenomenological fits to elastic scattering data.
In the future it is our intention to  compute these  effective  interactions  microscopically.

The CC method has been  shown to provide an accurate description of  low--lying spectra and properties of nuclei with closed (sub--)shells and their neighbors\cite{bartlett2007,hagen2014,ccemerg,ccni}.
We will employ the $\rm{NNLO_{sat}}$ interaction \cite{n2losat} derived  from chiral effective field theory, which  provides an accurate description of masses and radii in a wide mass--range.
The latter feature is critical for our approach, since a proper reproduction of the
distribution of nuclear matter, and, more specifically, nuclear radii, are essential to give an
accurate account of reaction observables. 

The GFT 
allows for the computation of $(d,p)$ cross section for bound, resonant and continuum states in the $(A+n)$ nucleus
by making use of the Green's Function of the $n-A$ subsystem \cite{Potel:15b,Potel:17b}. Within this context, the use of the prior form, in which the $n-A$ interaction appears explicitly in the $T$-matrix, is a way to avoid well known numerical difficulties in the post form associated with the convergence of the $T$-matrix integral in the continuum (see e.g. \cite{Thompson:09}). The ability to describe the population of continuum states is one of the main differences with respect to other approaches making  use of microscopic overlaps in the post representation \cite{flav}.

To benchmark this new approach (which we will denote as CC-GFT), we apply it to $(d,p)$ reactions on the doubly magic stable Ca  isotopes $^{40}$Ca and $^{48}$Ca, for which there is abundant data.
We then make predictions for  $(d,p)$ reaction cross sections on the short-lived $^{52,54}$ Ca, given the  recent experimental interest. Precision measurements of nuclear masses for $^{52}$Ca \cite{wienholtz2013} and 2$^{+}$ excitation energies for $^{54}$Ca \cite{steppenbeck2013b} suggest that these nuclei are also doubly-magic.
   
   This paper is organized as follows. We begin by summarizing the CC-GFT approach.
  We then demonstrate the applicability and reliability of the method by benchmarking our calculations
  with experimental data measured on the stable isotopes $^{40,48}$Ca, and  proceed to make  predictions for the $(d,p)$ transfer cross sections on the exotic  Calcium isotopes $^{52,54}$Ca.
  
\section{Formalism} \label{formalism}
In this section, we present a short description of both the GFT and  CC methods.
The GFT framework is based on a reaction formalism  introduced in the 1980's to address reactions
in which a fragment of the projectile fused with the target, while
the depleted projectile is detected with energies and angular distributions characteristic of a direct process \cite{Austern:81,Udagawa:80,Ichimura:85}.
The formalism has been recently revived and extended by several groups  \cite{Potel:15b,Lei:15,Carlson:15,Lei:15b,Potel:17b}.
In the next paragraphs, we present a  summary of the formalism following the derivation in \cite{Potel:15b}.
We  define $|\chi_p\rangle$ ($|\chi_d\rangle$) as the proton (deuteron) elastic scattering solution   of the potential $U_p$ ($U_d$) at the energy $E_p$ ($E_d$).
$|\chi_p\rangle$ and $|\chi_d\rangle$ are  functions of, respectively, the $p-(A+1)$ and $d-A$ coordinates. Following standard practice in reaction theory, we compute the $(d,p)$ process as a perturbation of the dominant elastic channel, and write the incoming channel in a factorized form,
\begin{align}\label{eq100}
\ket{\Psi}\approx\ket{\chi_d}\ket{\phi_A}\ket{\phi_d},
\end{align}
where $\ket{\phi_A}$ is the ground state of the target $A$, and $\ket{\phi_d}$ designates the deuteron ground state.
The inclusive cross section is obtained by summing over all states $\ket{\phi_B^c}$ in the $B\equiv n+A$ system. In the \textit{prior} representation, 
\begin{align}\label{eq25}
 \frac{d^2\sigma}{d\Omega_pdE_p}=\frac{\mu_p\mu_dk_p}{4\hbar^4\pi^2k_d}\sum_c\left|\left\langle\chi_p\phi_B^c|V\,|\,\phi_d\phi_A\chi_d\right\rangle\right|^2\delta(E{-}E_p{-}\varepsilon_B^c),
\end{align}
where $\varepsilon_B^c$ is the energy of $\ket{\phi_B^c}$,  $\mu_p$ $( \mu_d)$ the proton (deuteron) reduced mass and $k_p$ $( k_d)$ the proton (deuteron) momentum. The transition potential $V$ in Eq.~\ref{eq25} is given by:
\begin{align}\label{eq82}
V=V_{n-A}+U_p-U_d,
\end{align}
where $V_{n-A}$ is a  neutron-target optical potential.
Let us introduce the following propagator in $B$ (see \cite{Potel:15b} for more details): 
 \begin{align}\label{eq26}
 G_{B}(E)=\lim_{\epsilon\rightarrow 0}\sum_c\frac{\ket{\phi_B^c}\bra{\phi_B^c}}{E-E_p-\varepsilon_B^c+i\epsilon}.
 \end{align}
Using the relation
 \begin{align}\label{eq105}
\text{Im}\,G_{B}(E)=-i\pi\,\delta(E{-}E_p{-}\varepsilon_B^c)\ket{\phi_B^c}\bra{\phi_B^c},
\end{align}
we can write Eq.~\ref{eq25} as
\begin{align}\label{eq104}
 \frac{d^2\sigma}{d\Omega_pdE_p}=-\frac{\mu_p\mu_dk_p}{4\hbar^4\pi^3k_d}\text{Im}\left\langle\,\phi_d\phi_A\chi_d|V^\dagger\,|\chi_p\right) G_B\left(\chi_p|V\,|\,\phi_d\phi_A\chi_d\right\rangle,
\end{align}
where   the round brackets indicate that we only integrate over the proton coordinate.
The formalism then allows  to address the population of both bound and continuum final states, and to disentangle 
the contribution from elastic  and non-elastic breakup  to the total proton singles (for details, see \cite{Potel:15b}).
The projection of $G_{B}(E)$ on the ground state $|\phi_A\rangle$ corresponds to single-particle Green's Function $G(\mathbf r,\mathbf r',E)$:
\begin{align}\label{eq101}
G(\mathbf r,\mathbf r',E)=\braket{\mathbf r,\phi_A|G_{B}(E)|\phi_A,\mathbf r'},
\end{align}
 where  $\mathbf r$ ($\mathbf r'$) is the relative $n$--$A$ coordinate.
 After some formal manipulations \footnote{The presence of the potential $V$ makes the projection operation  not trivial to perform, and the non-commutativity of $V$ and $|\phi_A\rangle$  (more precisely, the fact that $[V,\ket{\phi_A}\bra{\phi_A}]\neq0$) has to be taken into account (see \cite{Potel:15b}).} , the cross section can be expressed in terms of $G(\mathbf r,\mathbf r',E)$ and a non-orthogonality contribution
 $\psi_n^{HM}=\left(\chi_p\right|\,\left.\,\vphantom{\chi_f}\phi_d\,\chi_d\right\rangle$ (also called Hussein-McVoy term \cite{Hussein:85}).
The final expression for the non-elastic breakup contribution is,
\begin{align}\label{eq103}
\frac{d^2\sigma}{d\Omega_pdE_p}  =-\frac{\mu_p\mu_dk_p}{4\hbar^4\pi^3k_d}\left[\text{Im}\langle\,\rho|G\,|\rho\rangle  +2\text{Re}\langle\,\psi_n^{{HM}}|W_{n-A}\,G\,|\rho\rangle+ \langle\,\psi_n^{{HM}}|W_{n-A}\,|\psi_n^{{HM}}\rangle\right],
\end{align} 
where $W_{n-A}$ is the imaginary part of $V_{n-A}$, and  we have introduced the breakup density amplitude (called $S_{prior}$ in ref. \cite{Potel:15b})
\begin{align}\label{eq2}
\rho(\mathbf r,E)=\left(\chi_p|V_{n-A}(\mathbf r,\mathbf r',E)+U_p-U_d|\phi_d\chi_d\right\rangle.
\end{align}
In this paper, we are interested in the population of the  ground state of the $A+1$ system
\footnote{Note that in the case of population of a bound state for the final nucleus, the GFT formalism is equivalent to the one-step DWBA
  approach \cite{Potel:15b}.}.
In that case, the imaginary part $W_{n-A}$ of
 the optical potential vanishes and the cross section is then given by:
\begin{align} \label{eq1}
 \frac{d^{2}\sigma}{d\Omega_pdE_p}=-\frac{\mu_p\mu_dk_p}{4\hbar^4\pi^3k_d}
\int\text{Im}\,G(\mathbf r,\mathbf r',E^{A+1}_{gs})\rho^*(\mathbf r,E^{A+1}_{gs})\rho(\mathbf r',E^{A+1}_{gs})\,d\mathbf r\,d\mathbf r',
\end{align}
where $E_{g.s.}^{A+1}$ is the ground state energy of the $A+1$ system.
For a beam energy in the center of mass  $E^{cm}_d$,  the outgoing proton  energy is $E^{cm}_p=E^{cm}_{d}+E_{bd}-E_{g.s.}^{A+1}$ with $E_{bd}=-2.22$ MeV  the deuteron binding energy. The intrinsic deuteron state $|\phi_d\rangle$ is taken as the
solution of an $s$-wave Woods-Saxon potential  which
reproduces the radius and binding energy of the deuteron \cite{Potel:15b,Potel:17b}.
In this paper, the Green's Function $G(\mathbf r,\mathbf r',E^{A+1}_{gs})$ in Eq.~\ref{eq1}, the optical potential
$V_{n-A}(\mathbf r,\mathbf r',E^{A+1}_{gs})$ in Eq.~\ref{eq2}, as well as $E^{A+1}_{gs}$,
are microscopically computed with the coupled-cluster method.
For all nuclei considered in this work, the ground state in $A$ has $0^+$ spin and parity, which implies that
$G$ and  $V_{n-A}$  conserve angular momentum, and therefore, only the component with the  spin and parity $J^\pi$ of the ground state of $A+1$ contribute to Eqs. (\ref{eq2}) and (\ref{eq1}).
\\
\\
In the following we describe the main steps involved in the calculation of the Green's Function and $n-A$ optical potential
 with the coupled-cluster approach. For a more detailed description, see \cite{gfccpap,gfccpap2,cc_review}.
We start with the computation of the ground state of the target $A$. Working in the laboratory coordinates, the (intrinsic) many--body Hamiltonian $H$  reads
\begin{eqnarray}
H=\sum^{A}_{i=1}\frac{\vec{p_i}^2}{2m}
-\frac{\vec{P}^2}{2mA} +\sum_{i<j} V_{ij} +\sum_{i<j<k} V_{ijk}, 
\label {hami}
\end{eqnarray}
where $\vec{p_i}$ is the momentum of the nucleon $i$ of mass $m$ in the laboratory  and $\vec{P}=\sum_{i=1}^{A}\vec{p_i}$
 the momentum associated with the center of
 mass motion. The terms $V_{ij}$ and $V_{ijk}$ are nucleon-nucleon $(NN)$ and three-nucleon forces (3NFs), respectively.
 The single-particle basis solution of
  the Hartee-Fock potential generated by $H$ (\ref{hami}) is a good starting point for coupled-cluster calculations.
Denoting $|\Phi_{0}\rangle$ the Hartree-Fock state, the ground state of the target
is represented as 
\begin{eqnarray}
|\Psi_{0}\rangle=e^T|\Phi_{0}\rangle \label{cc1} , 
\end{eqnarray}
where $T$ denotes the cluster operator
\begin{eqnarray}
T &=& T_1+T_2+\dots= \sum_{i,a}t_i^a a^{\dagger}_a a_i+\frac{1}{4}\sum_{ijab}t_{ij}^{ab}t_{ijab}a^{\dagger}_aa^{\dagger}_ba_ja_i +\ldots .
\label{t_cluster}
\end{eqnarray}
The operators $T_1$ and $T_2$ induce $1p-1h$ and $2p-2h$ excitations of the reference state, respectively.
Here, the single-particle states $i, j, ...$ refer to hole
states occupied in the reference state $|\Phi_0\rangle$ i while $a, b, ...$
denote valence states above the reference state.
In practice, the expansion (\ref{t_cluster}) is truncated. In the coupled
cluster with singles and doubles (CCSD), which we will consider in this work, all operators $T_i$
with $i >$ 2 are neglected. 
In that case, the ground-state energy and the amplitudes
$t_i^a, t_{ij}^{ab}$ are obtained by projecting the state (\ref{cc1}) on the
reference state and on all $1p$-$1h$ and $2p$-$2h$ configurations.
Correspondingly, the CCSD ground state is an eigenstate of the similarity-transformed Hamiltonian $\bar{H}=e^{-T}He^{T}$ in the space of $0p-0h$, $1p-1h$, $2p-2h$ configurations. The transformed Hamiltonian $\bar{H}$ is not Hermitian due to the fact that the operator $e^T$ is not unitary.
As a consequence, it has left- and right-eigenvectors which constitute a bi-orthogonal basis.
We denote $\langle \Phi_{0,L}|$ the left eigenvector for the ground state of $A$.
From the definition of the Green's Function \cite{dickhoff}, we can  write the matrix elements of the
coupled cluster Green's Function $G^{cc}$ as
\begin{eqnarray}
G^{CC}(\alpha,\beta,E) \equiv \langle \Phi_{0,L}|\overline{a_{\alpha}}\frac{1}{E-(\overline{H}-E^{A}_{gs})+i\eta}\overline{a^{\dagger}_{\beta}}|\Phi_{0}\rangle +\langle \Phi_{0,L}|\overline{a^{\dagger}_{\beta}}\frac{1}{E-(E^{A}_{gs}-\overline{H})-i\eta}\overline{a_{\alpha}}|\Phi_{0}\rangle .
\label{gfcc}
\end{eqnarray}
where $\eta \rightarrow 0$ by definition. The operators $\overline{a_{\alpha}}=e^{-T}a_{\alpha}e^T$ and
$\overline{a^{\dagger}_{\beta}}=e^{-T}a^{\dagger}_{\beta}e^T$ are the
similarity-transformed annihilation and creation operators,
respectively.

In this work, we calculate  the Green's function $G^{ccsd}(E)$
starting from a CCSD approximation for the target.
We then compute the particle (hole) part of the Green's Function within a $2p-1h$ ($2h-1p$) space. By construction,
  the poles of the particle part
    of $G^{ccsd}(E)$ correspond to the
  energy  $E=E^{A+1}$ of the  $A+1$ system, solutions  of the particle-attached equation-of-motion (PA-EOM)
  coupled-cluster method truncated at the $2p-1h$ excitation level \footnote{similarly the Green's Function  $G^{ccsd}(E)$
  has also poles at $E=E^{A-1}$ which correspond to the eigenstates solution of the A-1 system obtained with the particle-removed equation-of-motion (PR-EOM) ~\cite{gour2006,gfccpap,gfccpap2}. } ~\cite{gour2006}.
  The optical potential $V_{n-A}^{ccsd}(E)$
  is then obtained by  inversion of the Dyson Equation fulfilled by $G^{ccsd}(E)$ \cite{gfccpap,gfccpap2}.
  This potential is non--local and energy--dependent and,  for scattering energies, also complex,  the imaginary component
   accounting for the loss of flux due to absorption into channels other than the elastic channel.
    For $E<0$, the spectrum of $V_{n-A}^{ccsd}(E)$ is the discrete set of bound state energies of the $A+1$ nucleus,  $E=E_{n}^{A+1}$.
 $G^{ccsd}(E_{gs}^{A+1})$ and $V_{n-A}^{ccsd}(E_{gs}^{A+1})$ are  then used in the GFT equations (\ref{eq1}) and (\ref{eq2}) to compute the $(d,p)$ cross section for the population of the $A+1$ ground state.

A comment is in order here.
The CC calculations are performed
 in the laboratory coordinates \cite{gfccpap,gfccpap2}
 whereas the Green's function and the $n-A$ potentials appear in the GFT equations (\ref{eq1},\ref{eq2})
 as functions of the relative coordinate
 $\mathbf r= \mathbf r_{n}-\mathbf r_{A}$ ($\mathbf{r}_{n}$, $\mathbf {r}_{A}$ are the laboratory coordinates
 of the neutron and the center of mass of the target $A$, respectively).
 In Eq. (\ref{eq1}) and (\ref{eq2}),  both quantities are implicitly identified
  to the  CC outputs  $G^{ccsd}(E)$ and $V_{n-A}^{ccsd}(E)$  calculated in the laboratory frame.
This introduces a small error (estimated in the next section) in the computed $(d,p)$ cross section which is a decreasing function of the target mass $A$ \cite{rcjohn}.

\section{Results}
The CC calculations are performed with the same inputs and   model spaces as in  \cite{gfccpap2}.
We work in a mixed basis of single particle (s.p.) Hartree--Fock states expanded either within the harmonic oscillator shells
or the Berggren basis \cite{berggren1968,berggren1971}, depending on  the partial wave considered.
Working within the Berggren ensemble provides a natural extension of the CC formalism into the complex--energy plane
\cite{ccemerg,hagen2007d,hagen2010a,Nicc} and allows to compute (weakly) bound and unbound solutions of the coupled--cluster equations (see also \cite{michel2009,yanen,simin,halo18}
for the use of Berggren basis in the context of configuration--interaction approaches). 
The  g.s. in  $^{41}$Ca, $^{49}$Ca and the exotic  $^{53}$Ca, $^{55}$Ca are particle bound.
Solving the CC equations in the Berggren basis then ensures that the radial asymptotic behavior of the $n-A$ potential is
properly accounted for any value of the separation energy $E_{gs}^{A+1}$. 
The s.p. basis contains harmonic oscillator shells such that $2n + l \leq N_{max}$  along with
a discretized set of Berggren states. While we show results for different $N_{max}$, we  fix the  number   of discretized-Berggren shells at $N_{sh}=50$, known to be sufficient for convergence  \cite{gfccpap,gfccpap2}.
 The nuclear part of the Hamiltonian is given by
 the chiral--EFT interaction $\rm{NNLO_{sat}}$ which consists of a nucleon--nucleon ($NN$)  and  three--nucleon forces (3NFs)
 and has been shown to provide an accurate description of masses and radii in a wide mass--range, and in
 particular for $^{40}$Ca and $^{48}$Ca  ~\cite{n2losat,hagen2015,garciaruiz2016,lapoux2016,duguet2017}.
The  $\rm {NNLO_{sat}}$ interaction includes  two--body  and three--body interaction terms \cite{n2losat}.
In all calculations, the maximum number of
quanta allowed in the relative motion of two nucleons ($N_2$), and
three nucleons ($N_3$), are equal  to $N_{max}$, except for
the most extensive calculations considered here, where $N_2=14$ and $N_3=16$.
We use the normal-ordered two-body approximation for the three-nucleon force term, which has
been shown to work well in light- and medium mass nuclei \cite{hagen2007a,roth2012}.
 The optical  potentials $U_d$ and $U_p$ for $^{40,48}$Ca$(d,p)$ are taken from \cite{Brown:74}. By design,
 they reproduce  deuteron and proton elastic scattering on the  $^{40,48}$Ca targets.
 Since no experimental data for elastic scattering on the exotic $^{52}$Ca and $^{55}$Ca is available, the parameters
 for $U_d$ and $U_p$  are taken in these cases from global systematics \cite{Koning:03,Han:06}.
 
\begin{figure}
          \centerline{\includegraphics*[width=9cm,angle=0,scale=0.75]{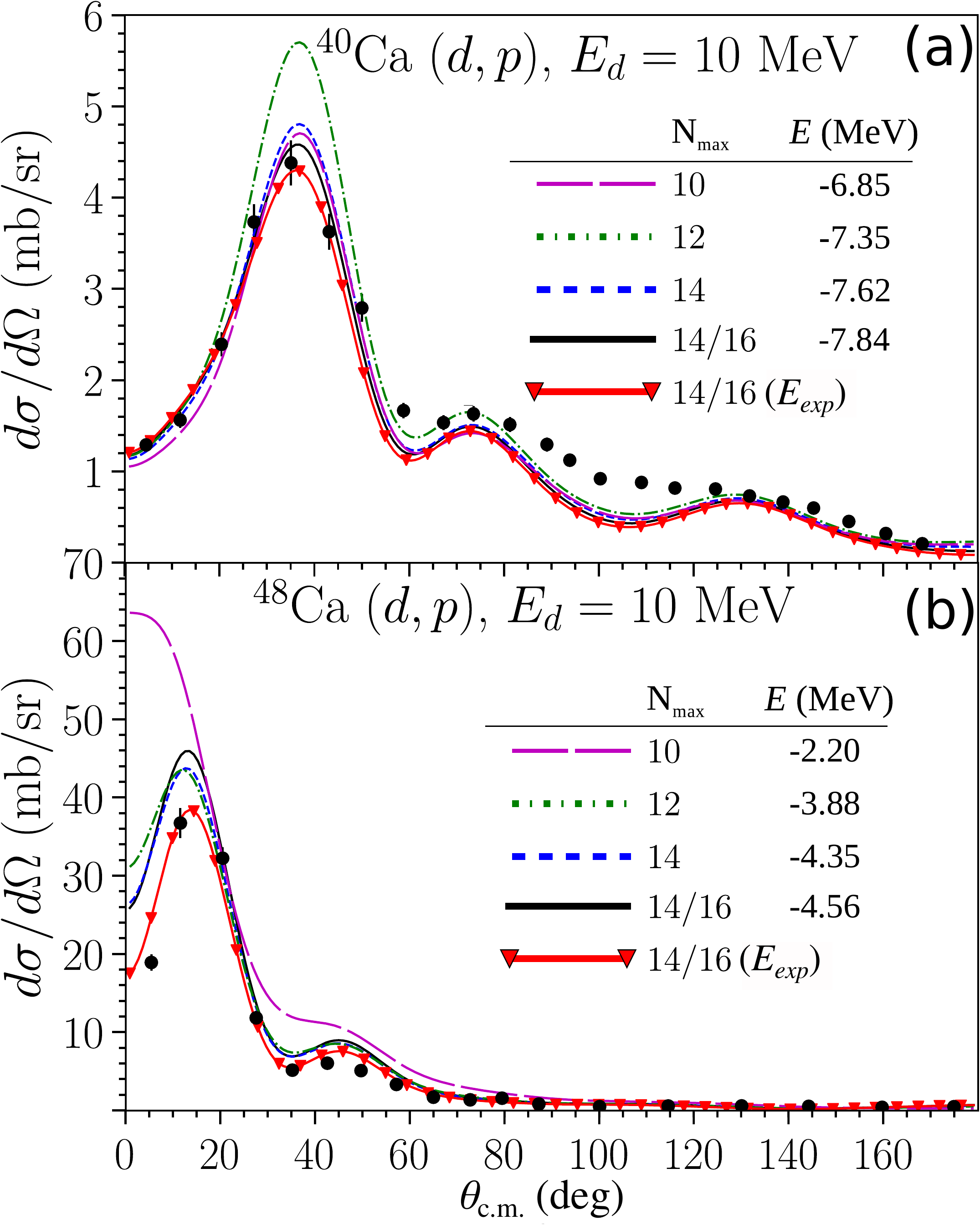}}
          \caption{ Calculations of the angular differential cross sections: a)   $^{40}$Ca$(d,p)^{41}$Ca(g.s.) at 10 MeV and b)   $^{48}$Ca$(d,p)^{49}$Ca(g.s.) at 10 MeV.  
The curves show the results of the CC-GFT calculations for different values of $N_{max}$. Also indicated are the PA-EOM energies $E_{gs}^{A+1}$.  The red  curve with triangles, labeled $N_{max}=14/16\,(E_{ex})$, was obtained within the largest model space by adjusting the energy to the experimental value. Theoretical calculations are compared with data (in full circle) from \cite{Brown:74}. }
\label{fig1}
	\end{figure}

The results for $^{40}$Ca$(d,p)^{41}$Ca and $^{48}$Ca$(d,p)^{49}$Ca at $E_d$=~10 MeV are shown in Fig.~\ref{fig1} 
as a function of the center of mass angle $\theta_{c.m.}$ for different values of $N_{max}$ along with the computed ground state energy in $^{41,49}$Ca. The converging pattern of the cross section is non--monotonic as $N_{max}$ increases,
and the calculated angular distributions for the largest model space {\it i.e.} $N_2/N_3=14/16$ are close to the data (see Fig.~\ref{fig1}).
We want to emphasize here that the CC computation of the inputs for the  few-body GFT equations have no free parameters.

\begin{figure}
	\includegraphics[width=8cm]{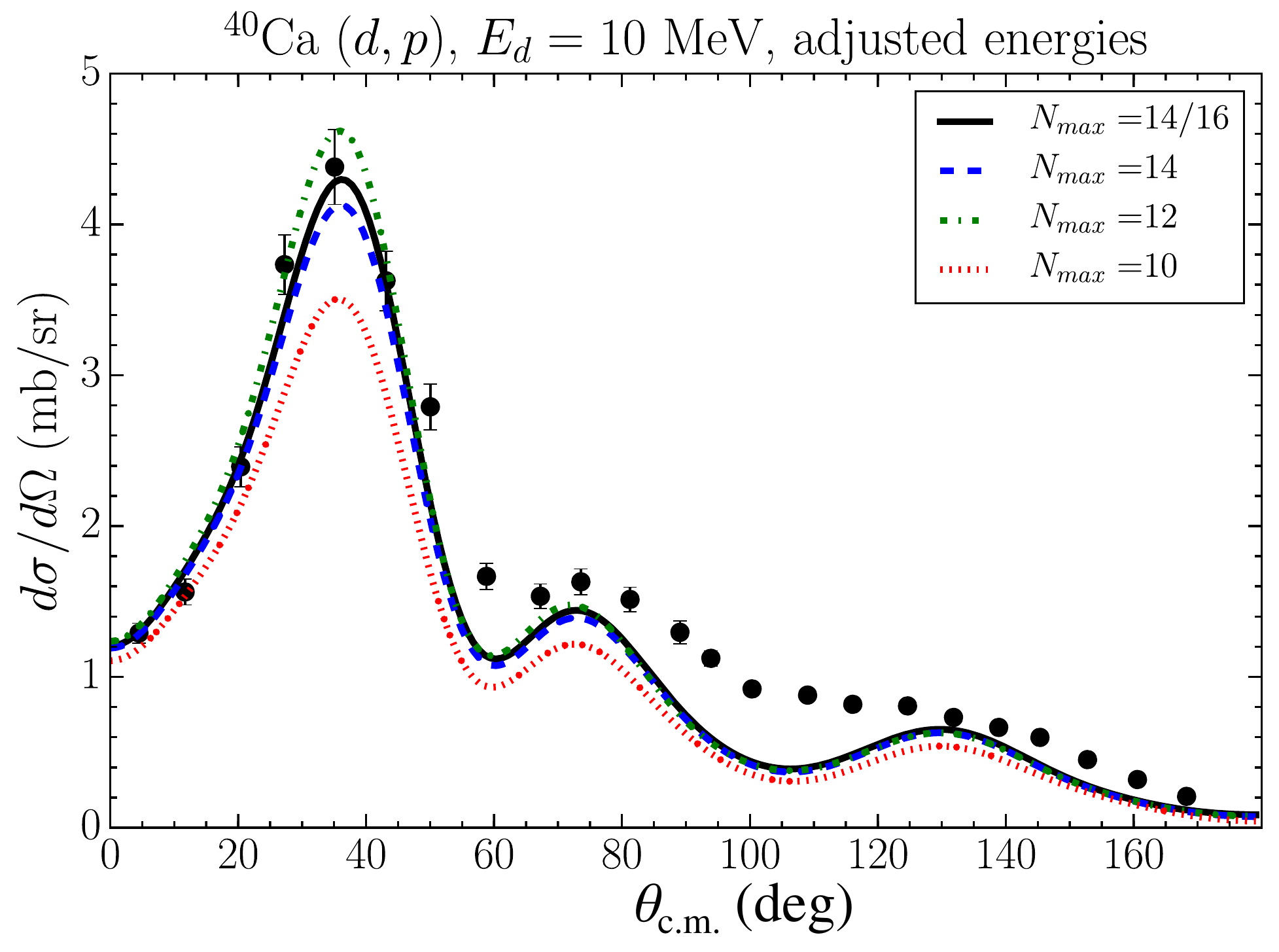}
	\caption{Angular distributions for the $^{40}$Ca($d,p$) reaction at different $N_{max}$ values. All calculations have been performed
		with the energy  $E^{A+1}_{gs}$ adjusted to the experimental value. }\label{figE}
\end{figure}

\begin{figure}
	\includegraphics[width=8cm]{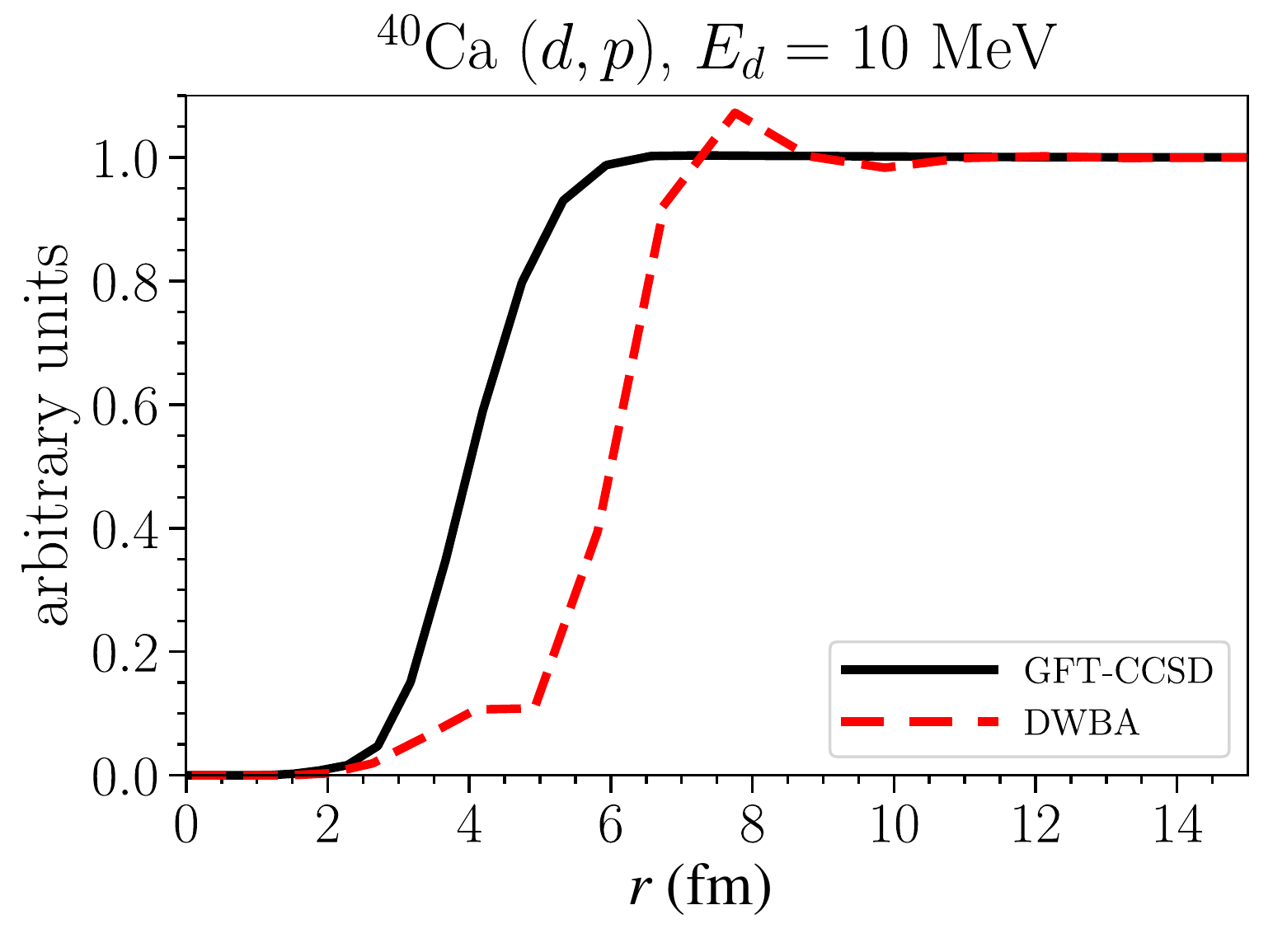}
	\caption{Cumulative contribution to the cross section as a function of the distance $r$ from the center of mass of $^{40}$Ca.
          For illustration, we show the results obtained in CC-GFT, $N_{max}=14/16$ and the results with DWBA, using with the same deuteron and proton optical potentials. The $n-A$ interaction and overlap function for the DWBA case have been obtained from a Woods-Saxon potential with standard values for the radius ($R_0=1.2A^{1/3}$ fm) and the diffusivity ($a=0.65$ fm).}\label{figR}
\end{figure}

For the largest model space, both nuclei are underbound at the PA-EOM level
by $\sim$ 500keV ($^{41}$Ca) and $\sim$ 600kev ($^{49}$Ca) with respect to
the experimental values   $E_{gs}^{A+1} = -8.36$ MeV ($^{41}$Ca) and $E_{gs}^{A+1} = -5.14$ MeV ($^{49}$Ca).
We can further improve the results by fixing the 
energy of the populated state to the experimental data while keeping all other inputs fixed.
In practise, this is done by adjusting the momentum $k_p$ of the proton in Eq. \ref{eq1}
and $\ket{\chi_p}$ in Eq. (\ref{eq2}), whereas
other quantities in the GFT equation  remain unchanged (we use $G^{ccsd}$ and $V_{n-A}^{ccsd}$ calculated at $N_2/N_3$=14/16).
We then obtain a remarkable agreement with the experimental data (red curve with triangles in Fig.~\ref{fig1}).
 We show in Fig.~\ref{figE} the calculated $^{40}$Ca$(d,p)^{41}$Ca(g.s.) cross section
  for each model space while fixing in all cases the ground state energy of $\rm{^{41}Ca}$ to the experimental data.
  The observed convergence is smoother than the one shown in Fig. \ref{fig1} and
   the computed cross section at the $\theta_{CM}\approx35^\circ$ peak corresponding to the $N=12$ calculation  is already within the experimental error bar.
   For illustration purpose, we show  in Fig.~\ref{figR} the cumulative contribution to the cross section (obtained by integrating over $r'$ in  Eq.~\ref{eq1}), as a function of $r$, both for CC-GFT and standard DWBA calculations. We want to emphasize here that the difference
   between both calculations stems from the $n-A$ inputs of the reaction equations: in the case of CC-GFT, the Green's Function and
   the optical potential are microscopically computed with the CC whereas for DWBA, the spectroscopic factor and the potential
   are fitted to reproduce the experimental transfer cross section. In that latter case, we use  a local potential of Woods-Saxon shape
   with standard values for the radius ($R_0=1.2A^{1/3}$ fm) and the diffusivity ($a=0.65$ fm). 
    The different contribution pattern seen in Fig.~\ref{figR} (the
          CC-GFT cross section gets contributions from deeper nuclear
     regions than the DWBA one) has its origin in the difference of the $n-A$ potentials.
   Keeping in mind that spectroscopic factors are scheme-dependent and only meaningfull within the context used to extract them,
   we obtain 0.897 and 0.895 for respectively $^{41}$Ca and $^{49}$Ca from the present CC calculation. 
  In the DWBA analysis of  \cite{Brown:74}, spectroscopic factors were extracted at two different beam energies:
  at $E_d=7$ MeV, $S=1.00$ and $S=1.002$  for $^{41}$Ca and $^{49}$Ca, respectively and at
  $E_d=10$ MeV, $S=0.850$ was obtained for $^{41}$Ca and $S=0.892$ for $^{49}$Ca.

Let us  mention here that previous GFT calculations  of the $^{40}$Ca$(d,p)^{41}$Ca cross section 
have  been performed in \cite{Potel:17b}, with inputs
from
the Dispersive Optical Model  \cite{DOM,waldecker} and that in \cite{flav},  post-form DWBA calculations with microscopic overlaps (computed with the self-consistent Green's function \cite{scgfa,scgfb})  have  been reported  for transfer reactions on Oxygen isotopes. 

As mentioned above, $U_d$ and $U_p$ are taken as phenomenological potentials fitted to reproduce
elastic scattering on $^{40,48}$Ca. Since these ``external'' interactions have been computed independently of
the $n-A$ potential, an uncertainty in the computed $(d,p)$ cross section will result.
Let us consider two interactions $U_{d1}$ and $U_{d2}$ that reproduce $d-A$ elastic scattering with the same quality.
At the two-body level, they are equivalent since by design they reproduce the data.
However, in the three-body system $(A,p,n)$, the differences in their off-shell behavior (which is not constrained by the fit)
will result into an uncertainty on the computed $(d,p)$ cross section.
In order to  estimate the uncertainty, we have performed calculations with  $U_d$ and $U_p$ 
fitted from global systematics  \cite{Koning:03,Han:06}.
We found a variation of less than 15\%  at the peak of the angular differential cross section, stemming mostly from the $d-A$ optical
potential. We should also point out here that the difference in the accuracy of the fits (a locally fitted interaction will certainly
reproduce the data more accurately than a global interaction) also contributes in this estimation.
In the future it is our intention to  compute these  effective  interactions  microscopically, consistently with the neutron-target potential.
Using, for instance, the Feshbach projection formalism \cite{Feshbach,Feshbach2}, $U_d$ could be derived from  the $n-A$, $p-A$ and $p-n$ potentials.


The difference between the center of mass coordinates used in the GFT equations (\ref{eq1}) and (\ref{eq2})
and the laboratory coordinates used in  CC introduces a small uncertainty in the calculation of the $(d,p)$
cross section, as discussed above. We estimate it by comparing the resulting difference in the cross sections when a shift $\Delta$ is added to
 $E_{gs}^{A+1}$ while keeping all other
inputs fixed in the GFT equations. We take $\Delta=E_{gs*}^A-E_{gs}^{A}$, where $E_{gs*}^{A}$ is the CCSD energy
for the ``mass-shifted'' nucleus $A$
\footnote{We recall that the PA-EOM calculation is a multistep procedure where one starts from
  the CCSD solution  (with energy $E_{gs*}^{A}$) of the mass-shifted nucleus $A$ as the reference state  to compute the energy
  $E_{*}^{A+1}$ in the $A+1$ system \cite{hagen2010a}.
  One  subsequently obtains  $E^{A+1}$ as $E^{A+1}=E_{*}^{A+1}+\Delta$, where the correction $\Delta=E^A_{gs*}-E_{gs}^{A}$
 amounts to a $1/A$ order effect. See e.g. \cite{hagen2010a,gfccpap} for more details.}.
This shift amounts to a $1/A$ effect, and for $^{40}$Ca, $\Delta= 190$ keV. This results in a small difference ($< 4\%$)
in the $^{40}$Ca$(d,p)^{41}$Ca cross section at the $\theta_{CM}\sim 40^\circ$ peak, smaller than the experimental  error bars (see Fig \ref{fig1}).

\begin{figure}
		\centerline{\includegraphics*[width=6cm,angle=0,scale=1.1]{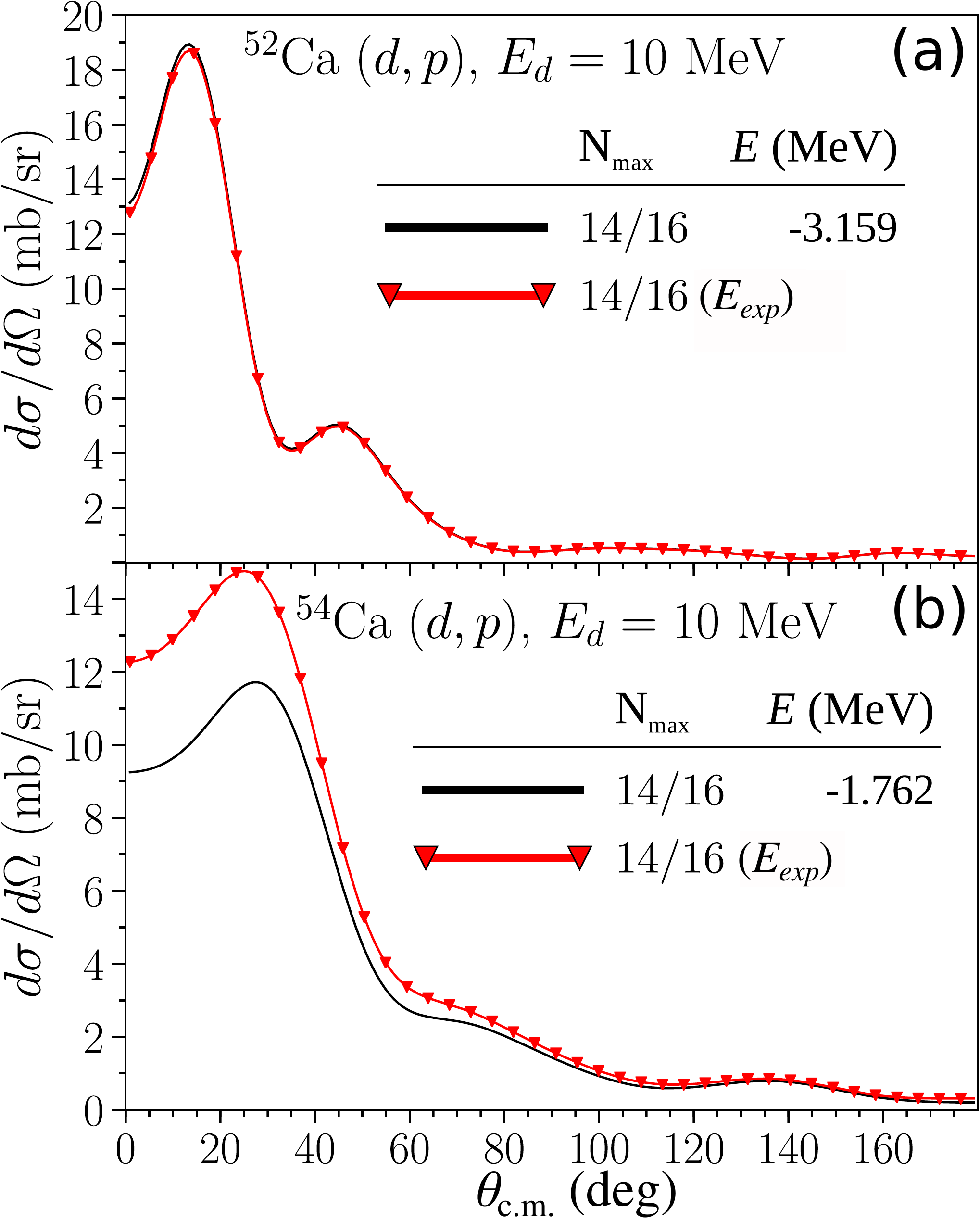}}
		\caption{ Predictions for the angular distributions: a)   $^{52}$Ca$(d,p)^{53}$Ca(g.s.) at 10 MeV and b)   $^{54}$Ca$(d,p)^{55}$Ca(g.s.) at 10 MeV.   We also list the energies  $E_{gs}^{A+1}$ calculated at the PA-EOM level to be compared with the experimental values $E=-3.46$ MeV for $^{53}$Ca and $E=-2.60$ MeV for $^{55}$Ca (see caption of Fig.~\ref{fig1} for more details). 
}
\label{fig2}
 \end{figure}

Encouraged with the good results obtained on the stable Ca isotopes, we  make predictions for the $(d, p)$ cross section with the unstable (although
particle--bound), neutron--rich $^{52}$Ca and $^{54}$Ca, for which experimental evidence of shell closure has been reported \cite{steppenbeck2013}.
Recent measurements have shown an increase in the charge radius of $^{52}$Ca (reproduced by CC calculations with $\rm{NNLO_{sat}}$)
with respect to what is expected for a double magic system \cite{garciaruiz2016}.
The required  beam intensity  for these experiments is expected to be achieved at FRIB from its first day of operation. 
The ground state energies for $^{53,55}$Ca are otherwise known ($E_{gs}^{A+1}=-3.46$ MeV for $^{53}$Ca and $E_{gs}^{A+1}=-2.60$ MeV for $^{55}$Ca), to be compared with the $N_2/N_3$=14/16  PA-EOM calculations ($E_{gs}^{A+1}=-3.16$ MeV for $^{53}$Ca and $E_{gs}^{A+1}=-1.76$ MeV for $^{55}$Ca).
The results are shown in Fig. \ref{fig2}. The  difference between the experimental and computed energies in $^{53}$Ca is $\sim$300 keV
whereas it is $\sim$ 820 keV for $^{53}$Ca.
This results in a larger difference between the  $^{54}$Ca$(d,p)^{55}$Ca  cross sections calculated with the PA-EOM energy
and the experimental value. 
%
%

        \section{Conclusion}
 We take in this paper an important step towards the development of a consistent microscopic theory for $(d,p)$ reactions
in medium-mass nuclei. Within a many-body framework where all nucleons are active, we  compute the Green's functions and $n$--$A$ optical potentials in the CC approach, with the two-and three-body $\rm{NNLO_{sat}}$  interaction. The $(d,p)$ cross section is then obtained by integrating the CC calculations in the GFT few-body formalism.
 We thus depart from   standard reaction formalisms in the following manner: in our approach, the observable cross section is reproduced from the consistent calculation, as enforced by the Dyson equation, of two
 non-observable quantities, namely the Green's function and the $n$--$A$ optical potential.
 Using phenomenological  $p-(A+1)$ and $d-A$ potentials,
 we obtain converged results in good agreement with available data for $^{40,48}$Ca, and show that the quality of the calculation can be improved further by adjusting the energy of the populated ground state to the experimental value.
 In the future, we plan to  compute these  effective  interactions  microscopically, en-par with the neutron-target input.
 CC calculations have been successful in reproducing the experimental findings regarding the exotic 
 isotopes  $^{52,54}$Ca, around the $N=32,34$ recently found closed shells. The formalism presented here allows for the integration of these CC calculations in the reaction framework, and 
can predict $(d,p)$ reaction cross sections for  $^{52,54}$Ca. These experiments are expected to be feasible in the near future at the new FRIB facility. 

 \section*{acknowledgments} 
We thank Gustav Jansen and Gaute Hagen for sharing the  complex Berggren basis  NN+3NFs Hartree-Fock and CCSD codes used in this work. 
This work was supported by the Office of Science, U.S. Department of Energy under Award Number DE-SC0013365, by the
U.S. Department of Energy, Office of Science, Office of Nuclear Physics, under the FRIB Theory Alliance award DE-SC0013617
and by the U.S. Department of Energy National Nuclear Security Administration under the Stewardship Science Academic Alliances program, NNSA Grant No. DE-FG52-08NA28552.
This work relied on iCER and the High Performance Computing Center at Michigan State University for computational resources.
\bibliography{refs}

\end{document}